\documentclass{PoS}

\title{Recent investigations of QCD at HERA}

\ShortTitle{Latest results from HERA}

\author{\speaker{Matthew Wing}\thanks{Also at DESY.}\\
        University College London\\
        E-mail: \email{m.wing@ucl.ac.uk}}

\abstract{
The latest results from the H1 and ZEUS collaborations which challenge the QCD description of high energy $ep$ collisions 
are presented.  Data from HERA continue to provide precision measurements and are compared to the latest theoretical 
predictions.  Measurements of new processes are also presented as well as investigation of regions where perturbative QCD 
fails to describe the data.  Four themes are presented here.  Measurements of hard QCD processes, prompt photon and jet 
production, are used to compare to the latest theoretical predictions and, in the case of jet production, used to make high-precision 
extractions of the strong coupling constant up to next-next-to-leading order in QCD.  All H1 and ZEUS charm and beauty cross 
sections in deep inelastic scattering have been combined and used to extract heavy-quark masses, including the running of the 
charm-quark mass with the scale of the process.  Factorisation in diffraction has been investigated in charm production in deep inelastic 
scattering and prompt photon production in diffractive photoproduction has been measured for the first time.  Finally, the inclusive 
data on deep inelastic scattering is presented in various forms in order to allow investigation of the underlying mechanism at 
very low photon virtuality $Q^2$ and low Bjorken $x$.
}

\FullConference{XXV International Workshop on Deep-Inelastic Scattering and Related Subjects\\
		3-7 April 2017\\
		University of Birmingham, UK}

\newcommand{\pom}{\mbox{\scriptsize\it I\hspace{-0.5ex}P}}

\begin{document}

\section{Introduction}

The HERA accelerator is still to date the world's only electron--proton collider and provided high energy collisions recorded by the H1 and ZEUS 
collaborations.  As with all high energy colliders, a wide range of physics has been investigated, with searches for new high energy 
phenomena, such as supersymmetry, precision measurements of particle properties, and, in particular, a deeper understanding 
of the strong interaction and its description, embodied in QCD.  Given the collision of a point-like object with the hadron, the investigation of 
the structure of the proton has been central to the HERA programme, with extension in low Bjorken $x$ and high virtualities, $Q^2$, by several 
orders of magnitude compared to fixed-target experiments.

Although HERA finished data taking 10\,years ago, H1 and ZEUS have published over 130\,papers since then, many investigating various 
aspects of QCD.  Precision measurements continue to be performed which can be used to extract fundamental QCD parameters such as 
the strong coupling constant, $\alpha_s$.  New processes continue to be measured for the first time, challenging theoretical calculations.  
Recent precision and first-time measurements and their understanding in QCD will be discussed in these proceedings.

The nominal parameters for HERA were the collision of protons with energy, $E_p$, of 920\,GeV and electrons with energy, $E_e$, of 27.5\,GeV, 
yielding a centre-of-mass energy, $\sqrt{s}$, of 318\,GeV.  The collider ran during the period 1992--2007, with $E_p = 820$\, in the initial years and 
lower values of $E_p$ in 2007 in order to measure the longitudinal structure function, $F_{\rm L}$.  Each of the two experiments, collected about 
500\,pb$^{-1}$ of data, with some measurements based on a combined sample of 1\,fb$^{-1}$.  Both H1 and ZEUS had reasonably standard 
detectors for collider physics, although with more instrumentation in the direction of the proton beam.

\section{Prompt photon production in deep inelastic scattering}

Although the simplest process in which a high energy photon is produced in $ep$ collisions is a purely electromagnetic reaction, the measurement 
of the prompt photon production cross section is sensitive to QCD through possible higher-order processes.  The measurements are also sensitive 
to the partonic densities in the proton or incoming photon as well as the "Pomeron", the object sometimes used to describe the colourless exchange 
in diffraction.  Prompt photon production within the Standard Model can often be a background to searches for new physics and so its production 
needs to be understood.

The measurement of the production of prompt photons is complementary to measurements of jet production (see Section~\ref{sec:jets}).  
Although the samples are much smaller, the study of prompt photons has a number of advantages: an isolated photon produces a clear 
experimental signature; it is produced in the hard scatter and is not subject to hadronisation; and there are fewer possible diagrams 
compared to parton--parton scattering.

\begin{figure}
\begin{center}
\includegraphics[width=0.9\textwidth]{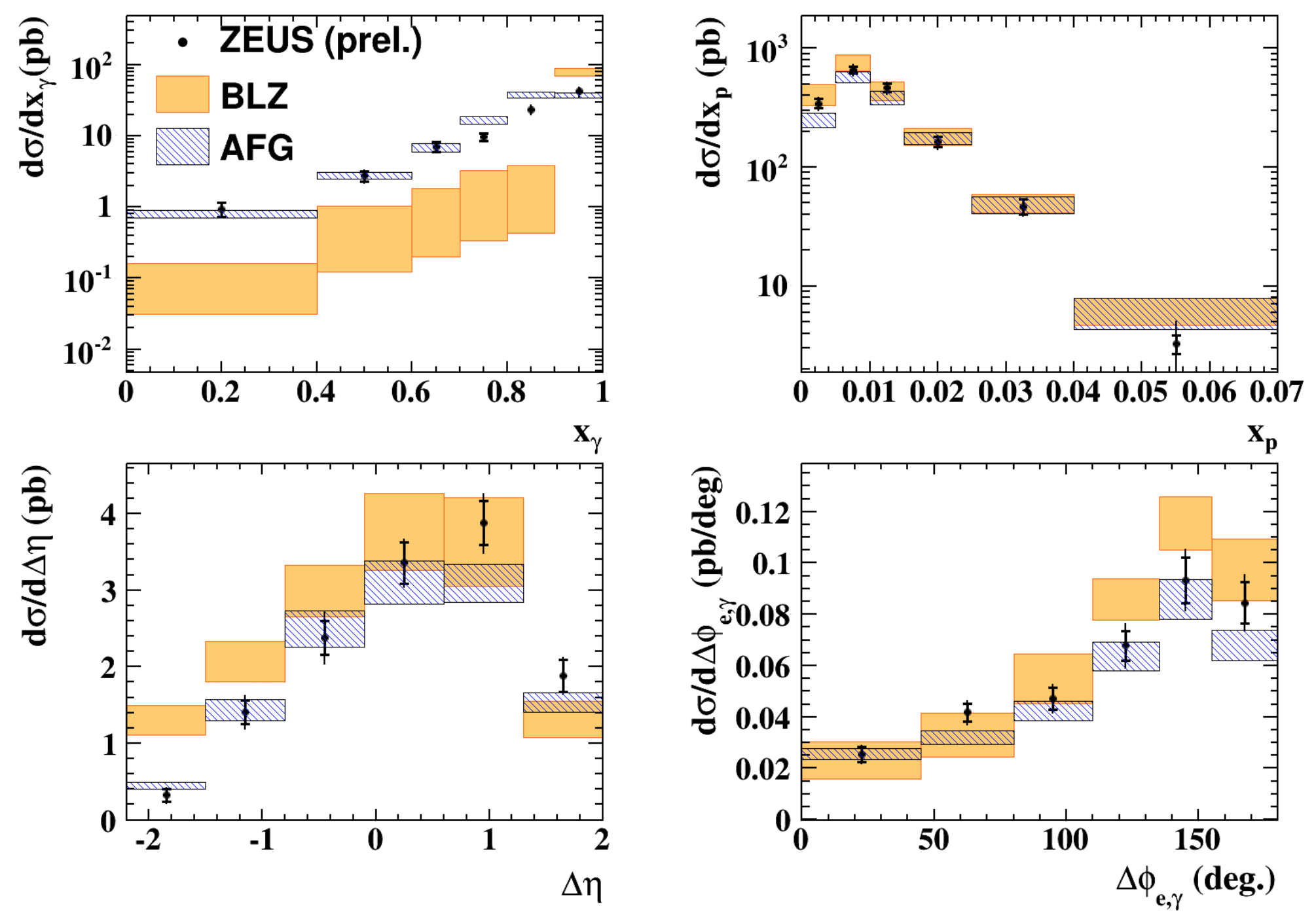}
\end{center}
\caption{Differential cross sections for various variables in prompt photon production in deep inelastic scattering.  The data are compared to 
theoretical predictions from Aurenche et al.\ (AFG) and from Baranov et al.\ (BLZ), both of which are corrected for hadronisation using the 
{\sc Pythia} Monte Carlo model.}
\label{fig:photon-dis}
\end{figure}

Prompt photon production with an accompanying hadronic jet in deep inelastic scattering~\cite{zeus-photons} is here presented in comparison to 
various models.  In Fig.~\ref{fig:photon-dis}, cross sections are shown as a function of the fraction of the incoming photon's momentum that takes 
part in the production of the photon and jet, $x_\gamma$, the fraction of the proton's momentum that takes part in the production of the photon and 
jet, $x_p$, the pseudorapidity difference between the jet and photon, $\Delta \eta$, and the azimuthal difference between the scattered electron and 
photon, $\Delta \phi_{e, \gamma}$.  The data are compared with two theoretical predictions: a next-to-leading order (NLO) QCD calculation by 
Aurenche, Fontannaz and Guillet~\cite{afg}, based on collinear factorisation; and a calculation by Baranov, Lipatov and Zotov~\cite{blz}, based on 
$k_T$ factorisation.  The prediction from BLZ does not describe the data in $x_\gamma$ and $\Delta \eta$, but gives a good description of the other 
variables (including others not shown).  The AFG prediction describes the data reasonably well.  When normalised to the data, the 
{\sc Pythia}~\cite{pythia} Monte Carlo simulation gives a good description (not shown here) of the data.

\section{Jet production in deep inelastic scattering}
\label{sec:jets}

\begin{figure}
\begin{center}
\includegraphics[width=0.8\textwidth]{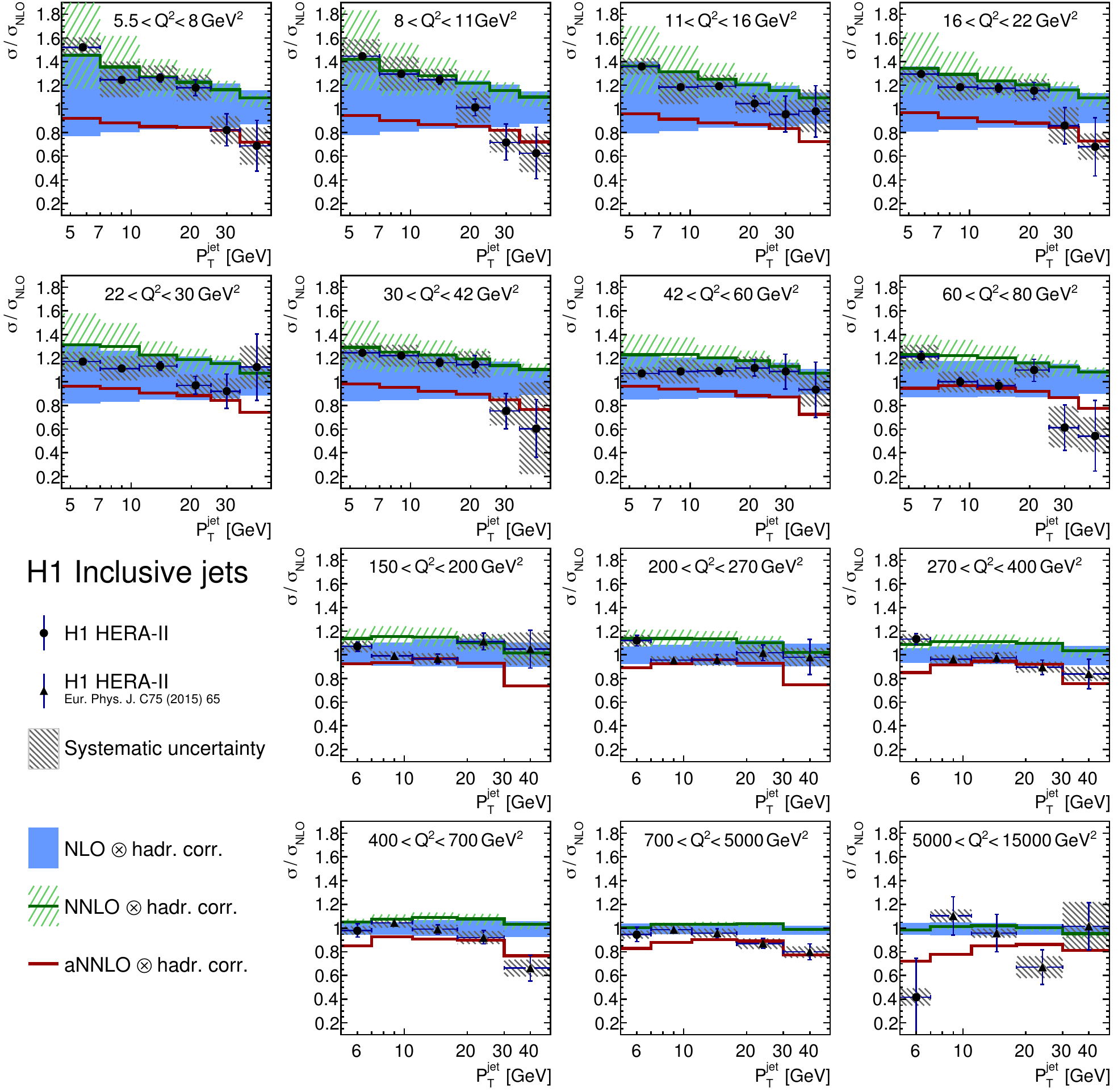}
\end{center}
\caption{Ratio of inclusive-jet cross sections to NLO QCD predictions and ratio of aNNLO and NNLO QCD to NLO QCD predictions as a function 
of $Q^2$ and jet transverse momentum, $P_T^{\rm jet}$.}
\label{fig:jets-dis}
\end{figure}

Jet production in deep inelastic scattering at HERA provides one of the most powerful tests of QCD, with sensitivity to the PDFs and precise 
extraction of the strong coupling constant, $\alpha_s$, possible.  The H1 collaboration have recently published such a measurement~\cite{h1-jets} using 
about 300\,pb$^{-1}$ and compared the data to pQCD calculations up to next-to-next-to-leading order (NNLO)~\cite{nnlo}.  The measurement 
was performed over a wide kinematic range, $5.5 < Q^2 < 15000$\,GeV$^2$; in Fig.~\ref{fig:jets-dis} cross sections are shown in comparison to 
different theoretical predictions.  The NLO QCD predictions describe the data reasonably well, although the theoretical uncertainties are relatively 
large, particularly at low $Q^2$.  The NNLO QCD predictions give an improved description and have smaller uncertainties due to the 
variation in the scales.

The data and theory have been used to extract values of $\alpha_s$ at NLO and NNLO.  The extractions agree with other extractions and with the 
world average, although the value using NNLO QCD is a little lower at $\alpha_s(M_Z) = 0.1157 (6)_{\rm exp} (^{+31}_{-26})_{\rm theo}$.  As can be 
seen from the numbers, the experimental uncertainty is about 0.5\%, whereas the theoretical uncertainty, even at NNLO, is significantly large 
at about 2.5\%, although reduced from about 4\% at NLO. 

\section{Extraction of the heavy-quark masses}

Heavy-quark production at HERA is dominated by boson--gluon fusion and as such is directly sensitive to the gluon density in the proton.  Given 
the masses of the heavy quarks, perturbative QCD predictions are expected to be reliable, although along with $Q^2$ and the transverse momentum 
of the heavy quarks, there are many scales in the process which poses an extra challenge to QCD calculations.  

\begin{figure}
\begin{center}
\includegraphics[width=0.7\textwidth]{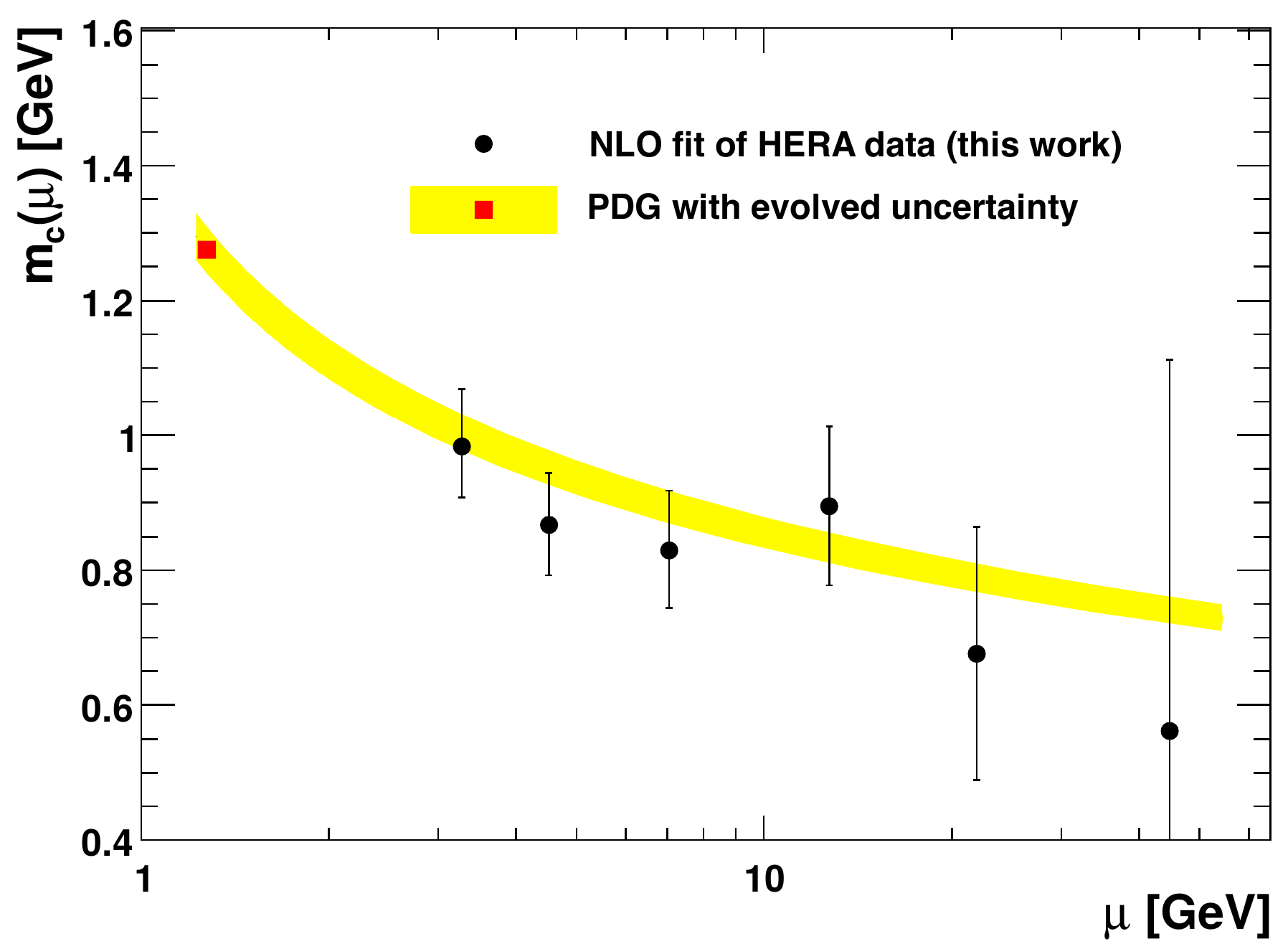}
\end{center}
\caption{Charm-quark mass, $m_c(\mu)$, as a function of the scale extracted from HERA data and compared to the PDG world average 
and its running evolved to higher scales.}
\label{fig:mc}
\end{figure}

Heavy-quark production in deep inelastic scattering is included in QCD fits to the parton densities in the proton and these analyses provide 
strong constraints on the masses of the heavy quarks.  Indeed the data can be used to extract the mass of the charm quark at several scales and 
investigate its evolution or "running" with scale~\cite{mc}, as is done for $\alpha_s$.  The mass as a function of the energy scale is shown in Fig.~\ref{fig:mc}, 
where the running of the charm-quark mass is clearly observed.

\begin{figure}
\begin{center}
\includegraphics[width=0.8\textwidth]{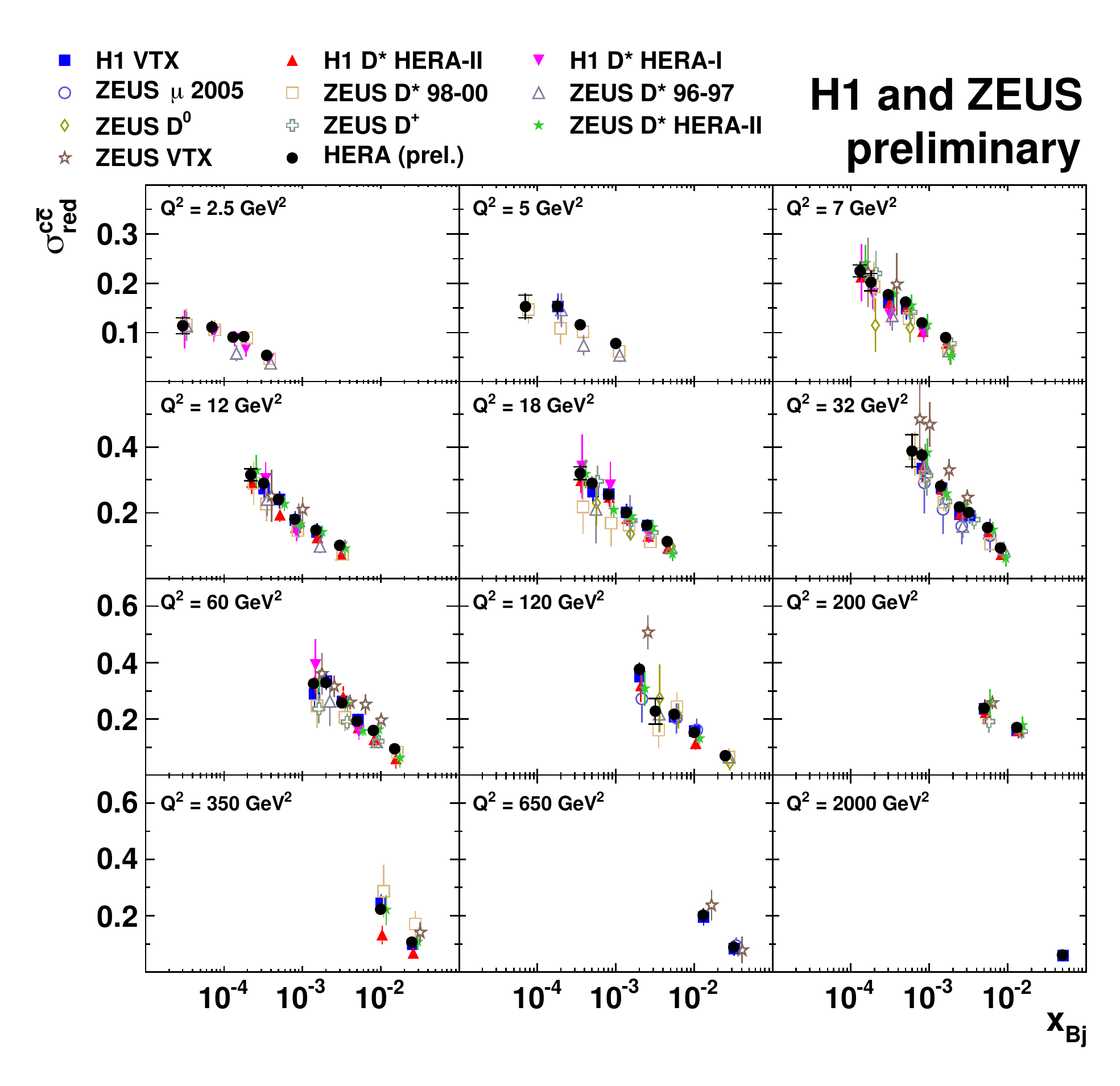}
\end{center}
\caption{Individual and combined cross sections for charm production in deep inelastic scattering.  The cross sections are shown as a 
function of Bjorken $x$ at fixed $Q^2$.}
\label{fig:c-comb}
\end{figure}

All of the charm data in deep inelastic scattering from H1 and ZEUS has been combined to form a common HERA data set~\cite{h1-zeus-b-c}, 
including three new high-precision sets compared to the previous combination~\cite{old-c-comb}.  The beauty data in deep inelastic scattering has 
been combined for the first time~\cite{h1-zeus-b-c}.  Correlations between all data sets have been accounted for.  The data sets are consistent 
($\chi^2/{\rm dof} = 149/187$) and the combination leads to an improved precision, as can be seen in Fig.~\ref{fig:c-comb}.

\begin{figure}
\begin{center}
\includegraphics[width=0.8\textwidth]{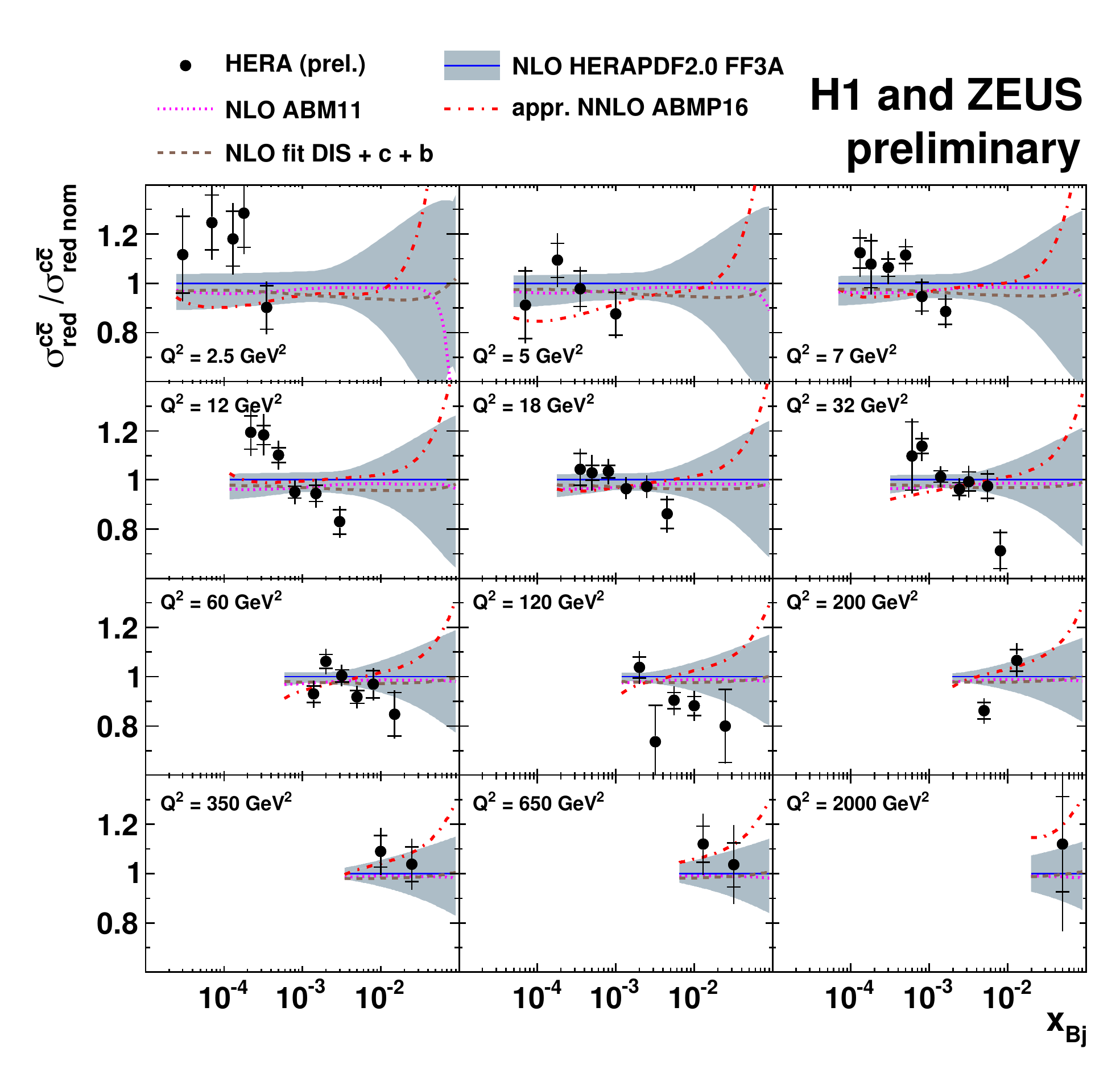}
\end{center}
\caption{Ratio of measured cross sections for charm production in deep inelastic scattering compared to QCD predictions with 
different proton PDFs: ABM11~\cite{abm11}, HERAPDF2.0 FF3A~\cite{hera-combined} and the fit discussed here all at NLO, as well as 
ABMP16~\cite{abmp16} at approximate NNLO.  The cross sections are shown as a function of Bjorken $x$ at fixed $Q^2$.  The grey bands 
represent the theory uncertainties.}
\label{fig:c-comb-ratio}
\end{figure}

The data are compared with NLO QCD predictions using different PDFs in Fig.~\ref{fig:c-comb-ratio}.  The description of the data is reasonable, 
although some difference in shape is seen, in particular for $Q^2 = 12$\,GeV$^2$.  The  NLO QCD analysis of the data ("NLO fit DIS + c + b" in 
Fig.~\ref{fig:c-comb-ratio}) yielded masses of the heavy quarks:

\begin{eqnarray}
m_c(m_c) & = & 1290^{+46}_{-41} ({\rm fit}) ^{+62}_{-14} ({\rm model}) ^{+7}_{-31} ({\rm param}) \, {\rm MeV}, \nonumber \\
m_b(m_b) & = & 4049^{+104}_{-109} ({\rm fit}) ^{+90}_{-32} ({\rm model}) ^{+1}_{-31} ({\rm param}) \, {\rm MeV}.
\end{eqnarray}

The results compare well with the PDG values ($m_c(m_c) = 1280 \pm 30$\,MeV and $m_b(m_b) = 4180^{+40}_{-30}$\,MeV)~\cite{pdg} and the charm mass in 
particular has competitive uncertainties.  The uncertainty on the extracted charm-quark mass presented here is dominated by variation of the scale in the 
QCD calculation.  It is hoped that improved, or higher-order, calculations will have an impact on the measurements, similar to that seen in the jet 
results (see Section~\ref{sec:jets}), and lead to an even more competitive extraction of the charm-quark mass.

\section{Diffraction and factorisation}

The observation and subsequent detailed investigation of diffraction in high energy $ep$ collisions is one of the highlights of the HERA 
programme.  The measurements from HERA re-invigorated this area of physics and still presents theory with a number of challenges such 
as the interplay between hard and soft physics, the question of factorisation and, in general, the nature of the underlying process.  The expectation of 
factorisation in diffractive deep inelastic scattering~\cite{collins} means that diffraction is amenable to QCD in which a hard sub-process is 
convoluted with the PDFs of the diffractive exchange, or "Pomeron", following a similar approach to that used for the proton in inclusive 
deep inelastic scattering.  

\begin{figure}
\begin{center}
\includegraphics[width=0.8\textwidth]{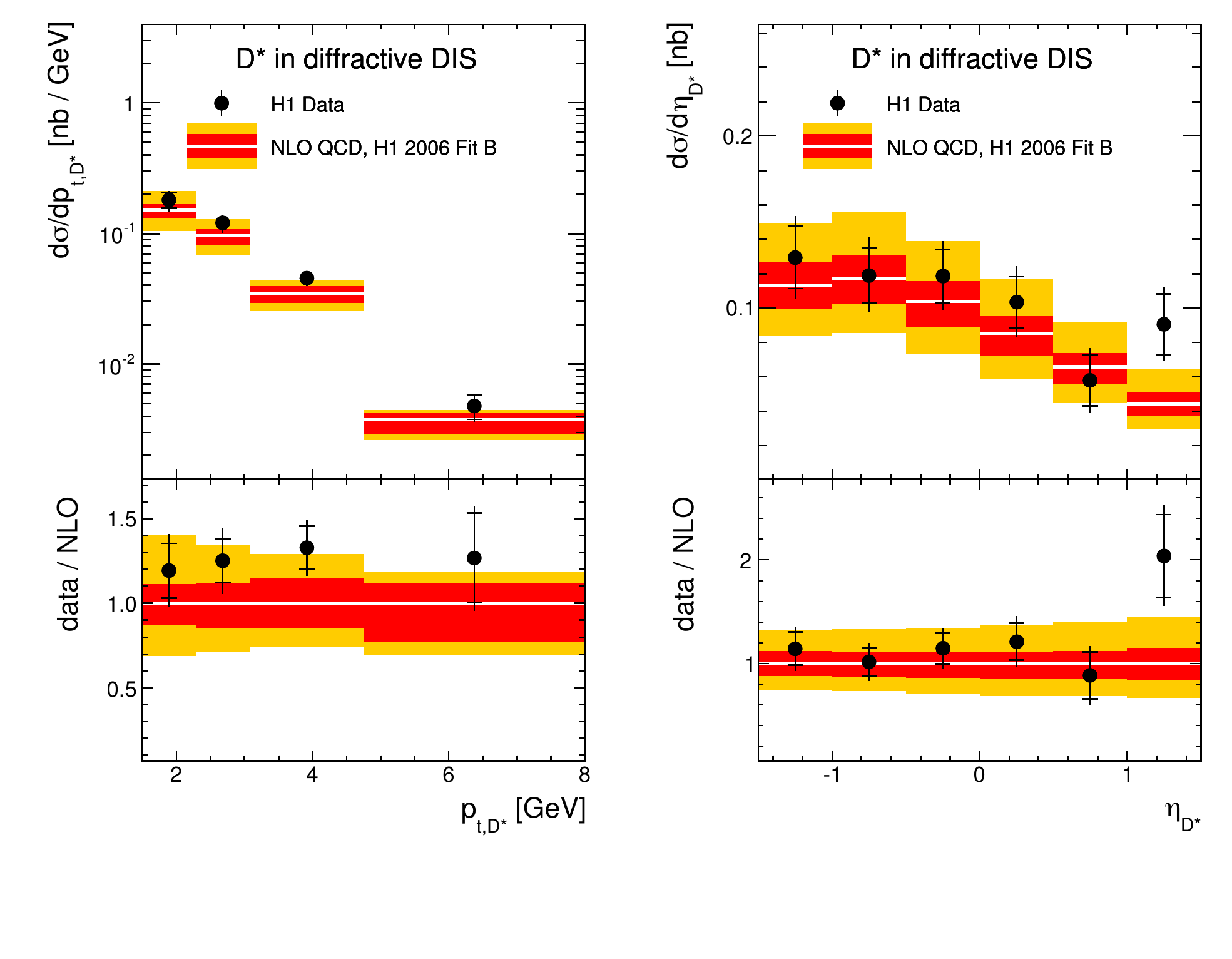}
\end{center}
\vspace{-1.5cm}
\caption{Cross sections for $D^*$ variables in diffractive deep inelastic scattering compared to predictions from NLO QCD.  The ratio of the 
data to NLO QCD predictions are also shown.}
\label{fig:charm-diff}
\end{figure}

As a result of factorisation, the diffractive PDFs extracted by fitting inclusive diffraction data should be applicable in other processes, such 
as when requiring a specific final state or having a different initial collision, e.g.\ $pp$.  In order to test this, the H1 collaboration have 
measured charm production, through reconstructing decays of $D^*$ mesons, in diffractive deep inelastic scattering using a luminosity of 
about 300\,pb$^{-1}$~\cite{h1-dstar-diff}.  The measurements are compared to predictions from NLO QCD with parton densities in the Pomeron extracted 
from fits~\cite{h1dpdf} to inclusive diffraction data in Fig.~\ref{fig:charm-diff}.  The variables, characterising the $D^*$ meson, {\em viz.} its transverse 
momentum, $p_{t, D^*}$, and pseudorapidity, $\eta_{D^*}$, are shown to be well described by QCD (as are other kinematic variables, not 
shown).  The conclusions are consistent with other results at HERA in which processes in diffractive deep inelastic scattering are well 
described by NLO QCD.

\begin{figure}
\begin{center}
\includegraphics[width=0.55\textwidth]{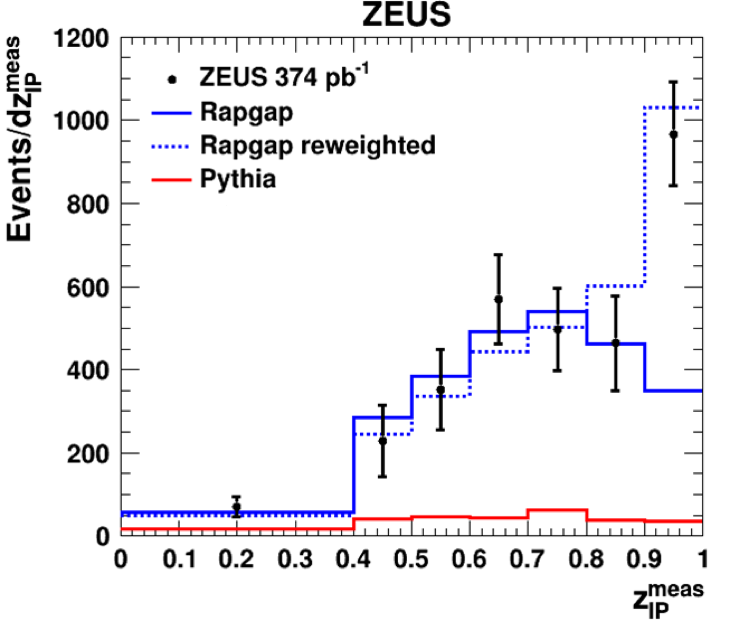}
\end{center}
\caption{Distribution of $z_{\pom}^{\rm meas}$ in ZEUS diffractive prompt photon photoproduction data compared 
to Monte Carlo simulations.  The default RAPGAP prediction is shown, as well as the prediction re-weighted to fit the $z_{\pom}^{\rm meas}$ 
distribution.  The prediction from PYTHIA shows the non-diffractive background contribution. }
\label{fig:photon-diff}
\end{figure}

In photoproduction, in which the exchanged photon is quasi-real ($\langle Q^2 \rangle \approx 10^{-4}$\,GeV$^2$), the holding of factorisation 
is not so clear.  Results from H1 indicate that NLO QCD cannot describe diffractive jet photoproduction, being roughly a factor of 2 too high, 
whereas the measurements from ZEUS indicate no such discrepancy~\cite{diff-jets}.  Therefore, it is imperative to measure other processes in 
photoproduction.  A first measurement~\cite{zeus-photon-diff} of diffractive prompt photon photoproduction has been made by ZEUS using a 
luminosity of 374\,pb$^{-1}$ and is shown compared to {\sc Rapgap}~\cite{rapgap} Monte Carlo expectations in Fig.~\ref{fig:photon-diff}.  Of particular note is 
the distribution of the fraction of the Pomeron energy that takes part in the production of a prompt photon and a jet, $z_{\pom}^{\rm meas}$,  
which is poorly described by the simulation.  The events at high $z_{\pom}^{\rm meas}$ are concentrated at high $x_\gamma$ and so are 
evidence for processes containing a direct photon and direct Pomeron, which is not present in {\sc Rapgap}.  Unfortunately, at the time of 
writing an NLO QCD prediction for this process is not available; it is hoped that a future calculation may shed more light on the situation of 
factorisation in diffractive photoproduction.

\section{Investigation of low-$Q^2$ data}

\begin{figure}
\begin{center}
\includegraphics[width=0.8\textwidth]{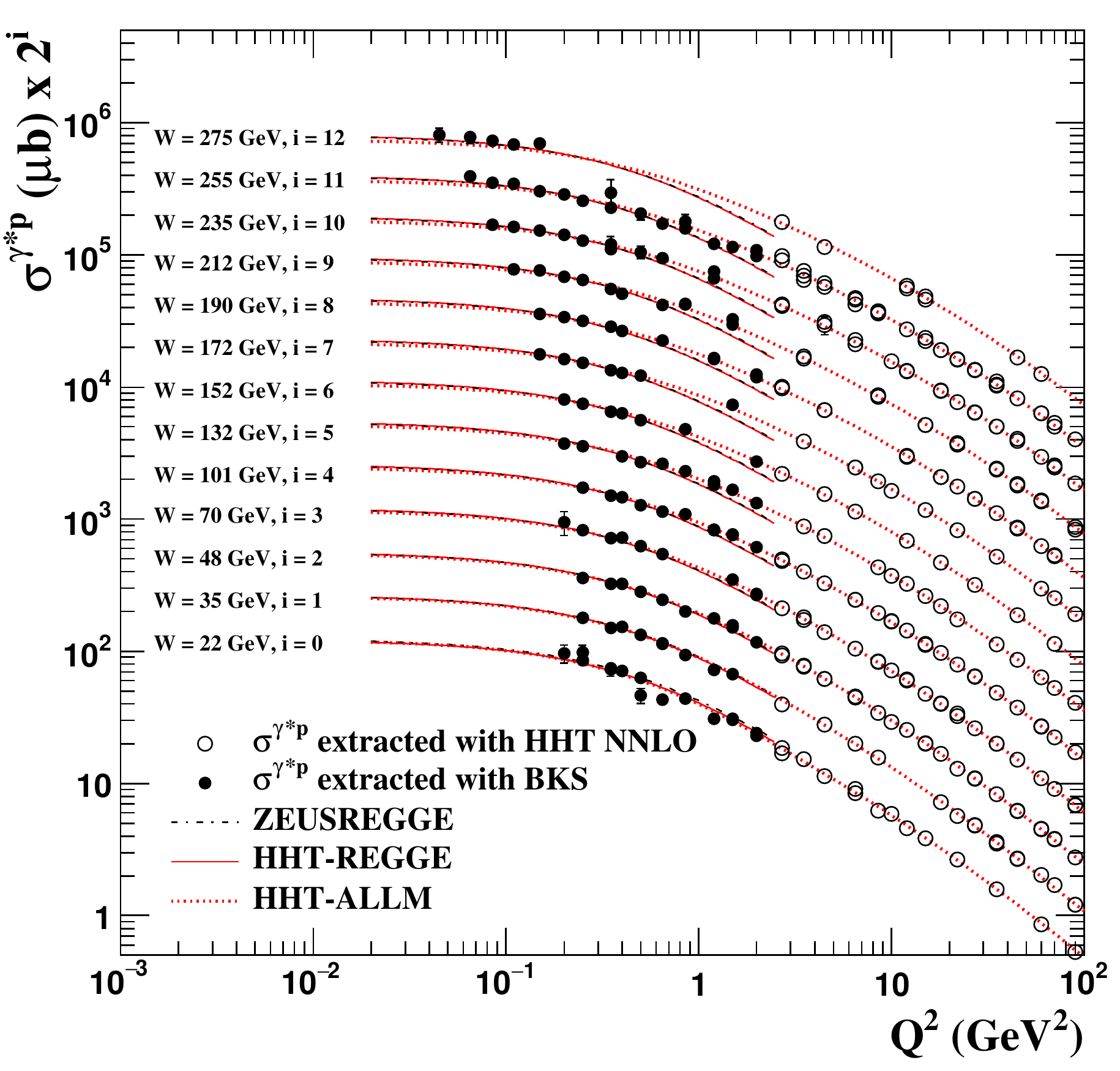}
\end{center}
\caption{Cross section, $\sigma^{\gamma^* p}$, versus $Q^2$  at fixed $W$, extracted with BKS at low $Q^2$ and with 
HHT NNLO QCD~\cite{hht1} at high $Q^2$.  The data are compared with a previous ZEUSREGGE fit as well as a new version, HHT-REGGE, 
and a new fit based on the ALLM model, HHT-ALLM.}
\label{fig:low-q2-w}
\end{figure}

The recently published final combined HERA data in inclusive deep inelastic scattering~\cite{hera-combined} provide a wealth of input to 
understanding QCD, such as being the pre-eminent data set in the extraction of proton PDFs and a precise extraction of $\alpha_s$.  The 
data can also be used to understand a region where pQCD is not expected to describe the data, at low $Q^2$ and low Bjorken 
$x$~\cite{lowq2}.  Data at the lowest $Q^2$ are shown in Fig.~\ref{fig:low-q2-w}, in which the virtual-photon--proton cross section is 
shown as a function of $Q^2$ for fixed photon--proton centre-of-mass energy, $W$.  The data show a smooth transition from the higher-$Q^2$ 
(perturbative) to the lower-$Q^2$ (non-perturbative) region.  These data cannot be described by pQCD predictions at NLO or NNLO, nor can 
they be described by a Regge model with one pole ("ZEUSREGGE"~\cite{zeusregge} and the updated "HHT-REGGE"~\cite{lowq2}).  The model 
HHT-ALLM~\cite{lowq2,allm}, 
which is also based on Regge phenomenology, but with many more parameters, describes the data reasonably well in the full kinematic region. 

\begin{figure}
\begin{center}
\includegraphics[width=0.8\textwidth]{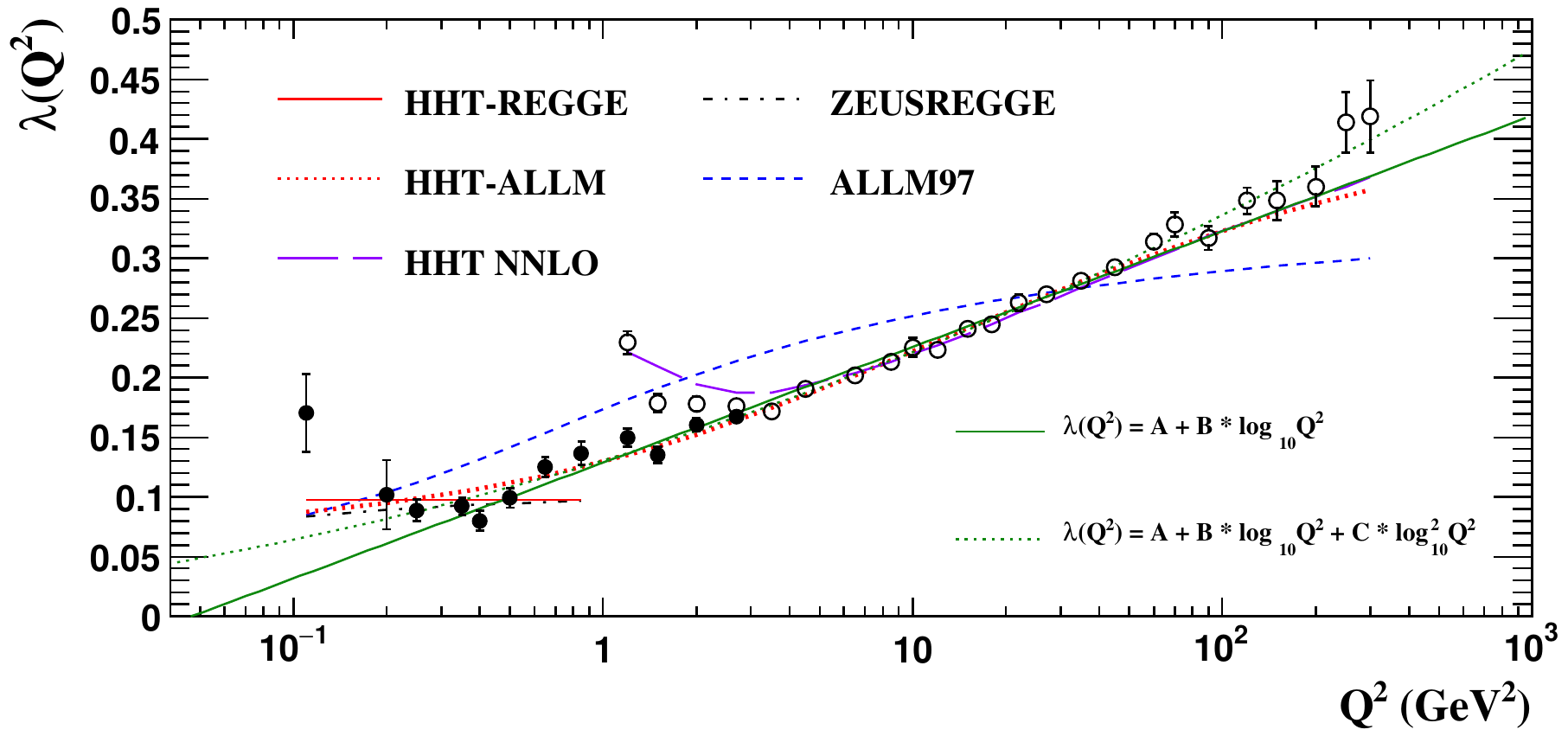}
\end{center}
\caption{Exponent $\lambda(Q^2)$ versus $Q^2$, from fits to $F_2$, extracted with NNLO QCD for $Q^2 \ge 1.2$\,GeV$^2$ (open points) and 
extracted with a Regge model for $Q^2 \le 2.7$\,GeV$^2$ (solid points).  The data are compared to various theoretical predictions, as well as a 
first- and second-order polynomial in $\log\,Q^2$.}
\label{fig:low-q2-lambda}
\end{figure}

The cross-section data can also be converted to the structure function, $F_2$, although this is a model-dependent procedure and was done 
with pQCD above a few GeV$^2$ and Regge phenomenology below.  The values of $F_2$ can then be parametrised as 
$F_2 = C(Q^2) x_{\rm Bj}^{-\lambda(Q^2)}$, where $C(Q^2)$ and $\lambda(Q^2)$ are parameters to be fit for each $Q^2$.  The extracted 
values of $\lambda(Q^2)$ are shown in Fig.~\ref{fig:low-q2-lambda}, compared to various expectations.  The low-$Q^2$ region can be 
described by simple Regge theory, whereas the high-$Q^2$ region can be described by pQCD and the re-fitted HHT-ALLM model 
describes the full-$Q^2$ region.  The overall trend with $Q^2$ is in fact rather simple and is well described by a quadratic fit.  These and 
other distributions will prove valuable in further theoretical development and understanding of the physics in this low-$Q^2$ and low-$x$ 
region.  The construction of a new $ep$ or $eA$ collider~\cite{ep-eA} would provide more data to widen the kinematic range of such future studies.

\section{Summary}

Results from HERA continue to challenge our understanding of QCD and, in these proceedings, some recent highlights have been reviewed.  
The data are compared to the latest predictions, up to next-to-next-to-leading order in QCD and this precision comparison allows world-competitive 
extractions of fundamental parameters of the strong force such as the strong coupling constant and masses of the heavy quarks.  Diffraction 
poses a special challenge although the validity of factorisation and use of diffractive parton densities is successful for deep inelastic scattering.  
However, a clear model to describe diffractive physics in the different initial states, deep inelastic scattering and photoproduction, and final states 
is still being sought.  Although perturbative QCD can describe many of the HERA measurements, in particular inclusive deep inelastic scattering, 
the region of low Bjorken $x$ is still poorly understood and further phenomenological development should be compared to the HERA data summarised 
here.

\section*{Acknowledgements}

The continuing efforts of the HERA collaborations and publication of high-quality results, 10\,years after the end of data taking, is 
truly remarkable.  These proceedings would not have been possible without this and as such are dedicated to the members of H1 and 
ZEUS who continue to produce results will challenge our understanding of QCD.  The support and hospitality from DESY, Hamburg 
is gratefully acknowledged as well as support from the Alexander von Humboldt Stiftung.

\end{document}